\documentclass[11pt]{article}
\usepackage[utf8]{inputenc}
\usepackage{booktabs}
\usepackage{hyperref}

\usepackage{graphics,graphicx}
\usepackage{times}
\usepackage{epstopdf}

\title{A Smart and Colorful Cadence for the \\ Wide-Fast Deep Survey }
\author{Suvi Gezari\footnote{suvi@astro.umd.edu}, (University of Maryland), \\
Sjoert van Velzen (University of Maryland and NYU), \\
Tiara Hung (University of California Santa Cruz), \\
Brad Cenko (NASA/Goddard), and \\
Iair Arcavi (Tel Aviv University)}

\date{November 2018}

\begin{document}

\maketitle

\begin{abstract}
Tidal disruption events (TDEs) are rare, 10$^{-7}$ yr$^{-1}$ Mpc$^{-3}$ (Hung et al. 2018), yet the large survey volume of LSST implies a very large detection rate of 200 yr$^{-1}$/1000 deg$^{2}$ (van Velzen et al. 2011), a factor of 250 increase in the  detection capability of the current generation of optical synoptic surveys, e.g. ZTF, ASAS-SN, Pan-STARRS, and ATLAS.  The goal of this LSST cadence white paper is to determine which survey strategy will ensure the efficient selection and characterization of TDEs in the LSST Wide-Fast-Deep Survey transient alert stream.  We conclude that the baseline cadence design fails to 1) measure the $u-r$ color and color evolution of transients, a critical parameter for distinguishing TDEs from SNe, and to 2) catch the pre-peak light curves of transients, an essential measurement for probing their rise times, which are in turn a probe of black hole mass in TDEs.  If we do not harvest the fruits of the LSST transient alert stream with photometric classification and early detections, both TDE and SN science will be greatly limited.  Hence, we propose for a ``smart" and "colorful" rolling cadence in the Wide-Fast Deep (WFD) Survey, that allows for efficient photometric transient classification from well sampled multi-band light curves, with the 20,000 deg$^{2}$ survey divided into eight 2500 deg$^{2}$ strips each observed for one year in Years 2-9, with the full WFD area observed in Years 1 \& 10.  This will yield a legacy sample of 200 TDEs per year with early detections in $u$, $g$, and $r$ bands for efficient classification and full light curve characterization.
\end{abstract}

\newpage
\section{White Paper Information}
\begin{enumerate} 
\item {\bf Science Category:} Exploring the Transient Optical Sky
\item {\bf Survey Type Category:} main `wide-fast-deep' survey
\item {\bf Observing Strategy Category:} 
    \begin{itemize} 
     \item a specific observing strategy to enable specific time domain science, 
	that is relatively agnostic to where the telescope is pointed (e.g., a science case enabled 
	by relatively deep precise time-resolved multi-color photometry). 
    \end{itemize}  
\end{enumerate}

\clearpage

\section{Scientific Motivation}

\begin{footnotesize}
 {\it Describe the scientific justification for this white paper in the context
of your field, as well as the importance to the general program of astronomy, 
including the relevance over the next decade. 
Describe other relevant data, and justify why LSST is the best facility for these observations.
(Limit: 2 pages + 1 page for figures.)}
\end{footnotesize}

Tidal disruption events (TDEs) provide a unique glimpse into the demographics, accretion physics, and environments of dormant supermassive black holes (SMBHs) lurking in the nuclei of galaxies.  Catching a TDE on its rise to peak, and monitoring its broadband and spectroscopic evolution can shed light on many important processes:\\
$\bullet$ the mass and spin of the central black hole\\
$\bullet$ the population and density profile of stars in the galaxy nucleus\\
$\bullet$ the efficiency of circularization of tidal debris into an accretion disk\\
$\bullet$ the production of a jet and the circumnuclear environment in which it interacts\\
$\bullet$ the presence of an outflow and/or unbound debris streams\\
$\bullet$ the time delay between the broadband emission components\\
$\bullet$ the detection of dust echoes and the inferred bolometric luminosity of the flare\\
$\bullet$ the kinematics, ionization state, and chemical composition of the tidal debris\\
$\bullet$ the detection of quasi-periodic oscillations indicative of radiation from the inner accretion flow\\
\indent While these results are exciting and illuminating, they have been possible for only a handful of TDEs with either well-sampled light curves, follow-up observations and detections across the electromagnetic spectrum, and/or multi-epoch spectra.  LSST has the survey volume to detect thousands of TDEs (Gezari et al. 2009; van Velzen et al. 2011), and yet 
the exciting science listed above is only possible if we can identify these sources against a background of SNe that is a factor $\sim$100 times larger. Given the high event rate and relatively faint magnitude (m$\sim$23 mag) of LSST transients, TDE/SNe identification will have to rely entirely on LSST photometry. {\bf Without sufficient multi-band coverage of the light curve, most TDEs and SNe will remain unidentified and thus "lost" forever in the LSST data. Furthermore, extracting the science listed above from the identified TDEs requires detections prior to the peak of the light curve. We thus arrive at two requirements that we will explore in this white paper: sufficient multi-band coverage for identification and a sufficiently high cadence observations to allow science with the identified sources.}

\begin{figure}[htp!]
\begin{center}
\hbox{
\includegraphics[scale=0.9]{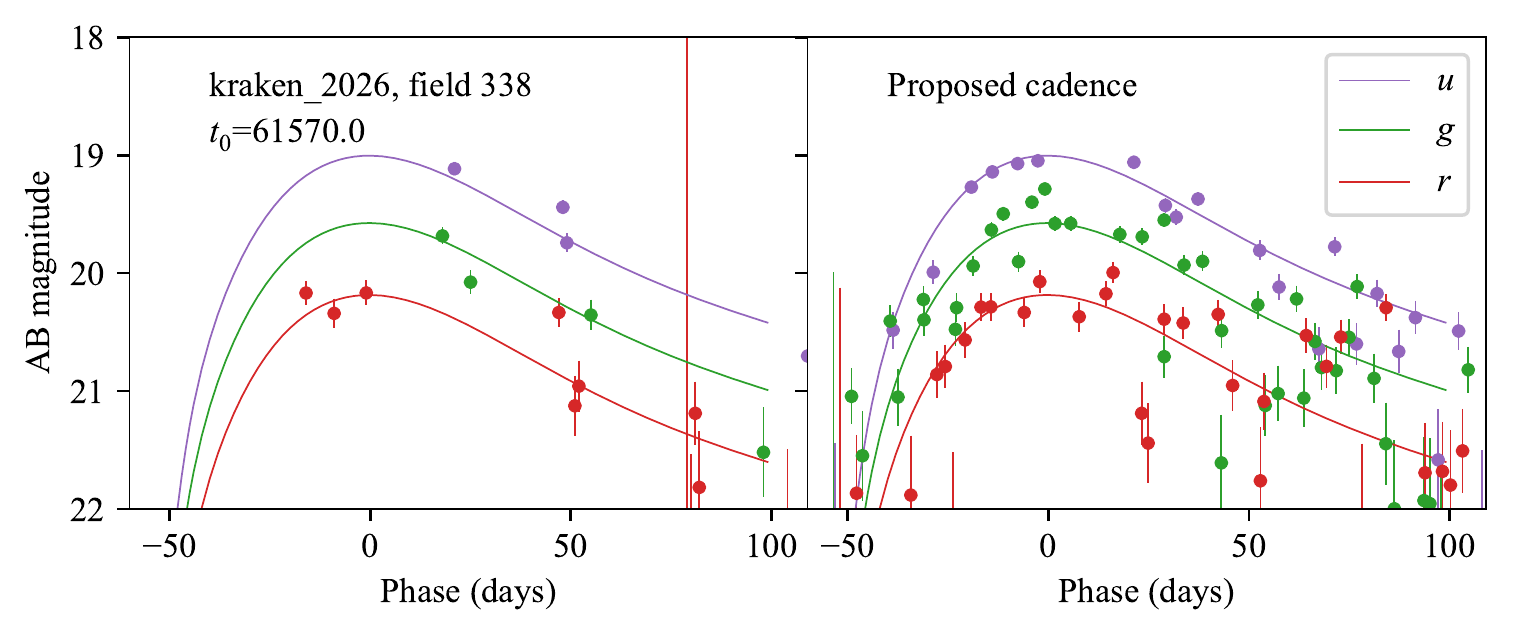}
}
\end{center}
\caption{
{\footnotesize {\ Left panel}: Simulated light curve of prototypical TDE PS1-10jh (Gezari et al. 2012), downsampled to the baseline (kraken\_2026) cadence.
{\it Right panel}: Simulated light curve at the cadence proposed in this white paper.  Note the dramatic improvement in catching the rise time to peak of the transient, and measuring its color and its color evolution both pre and post peak.}}
\label{fig1a}
\end{figure}

\begin{figure}[htp!]
\begin{center}
\hbox{
\includegraphics[scale=0.4]{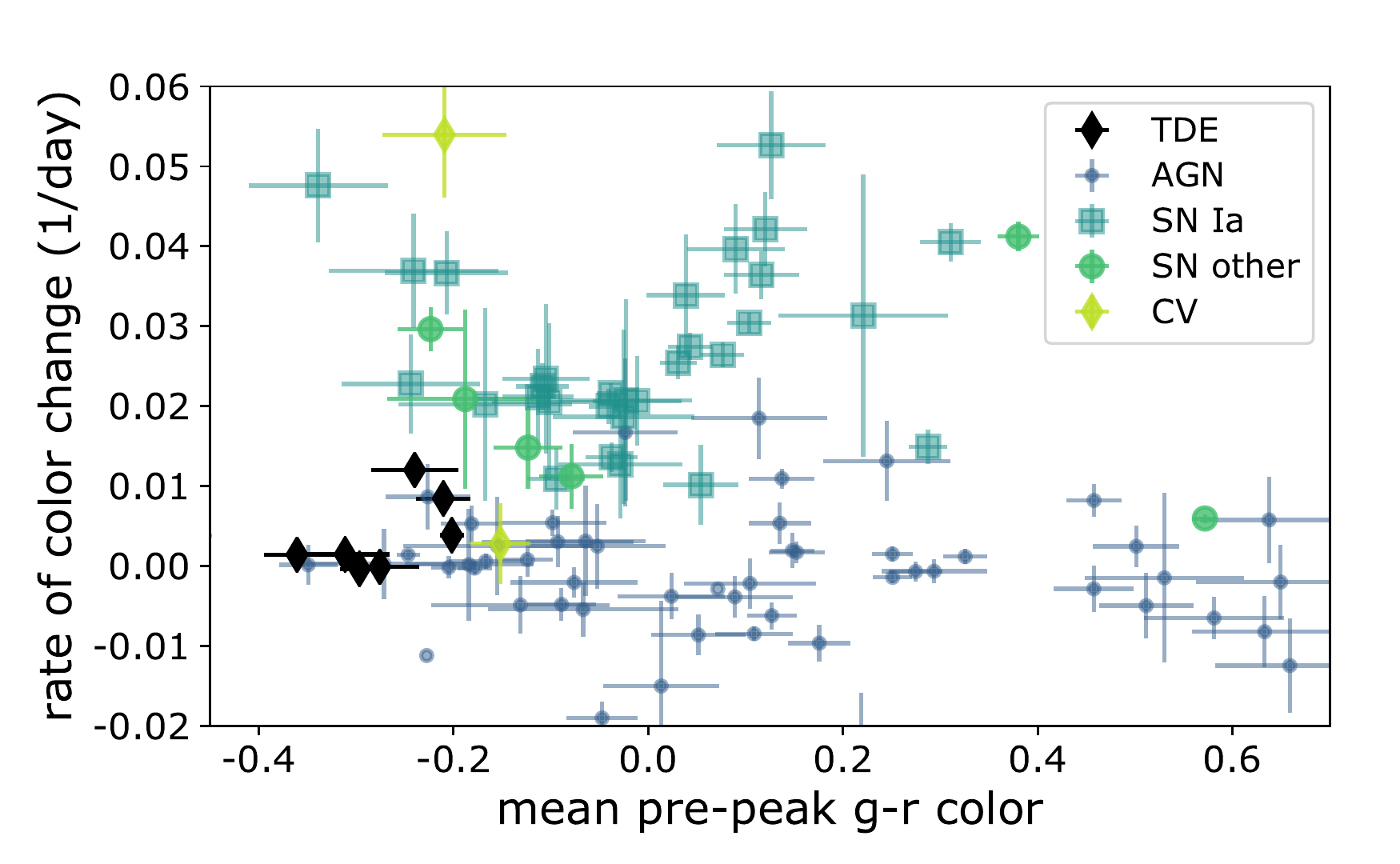}
\hspace{0.5cm}
\includegraphics[scale=0.4]{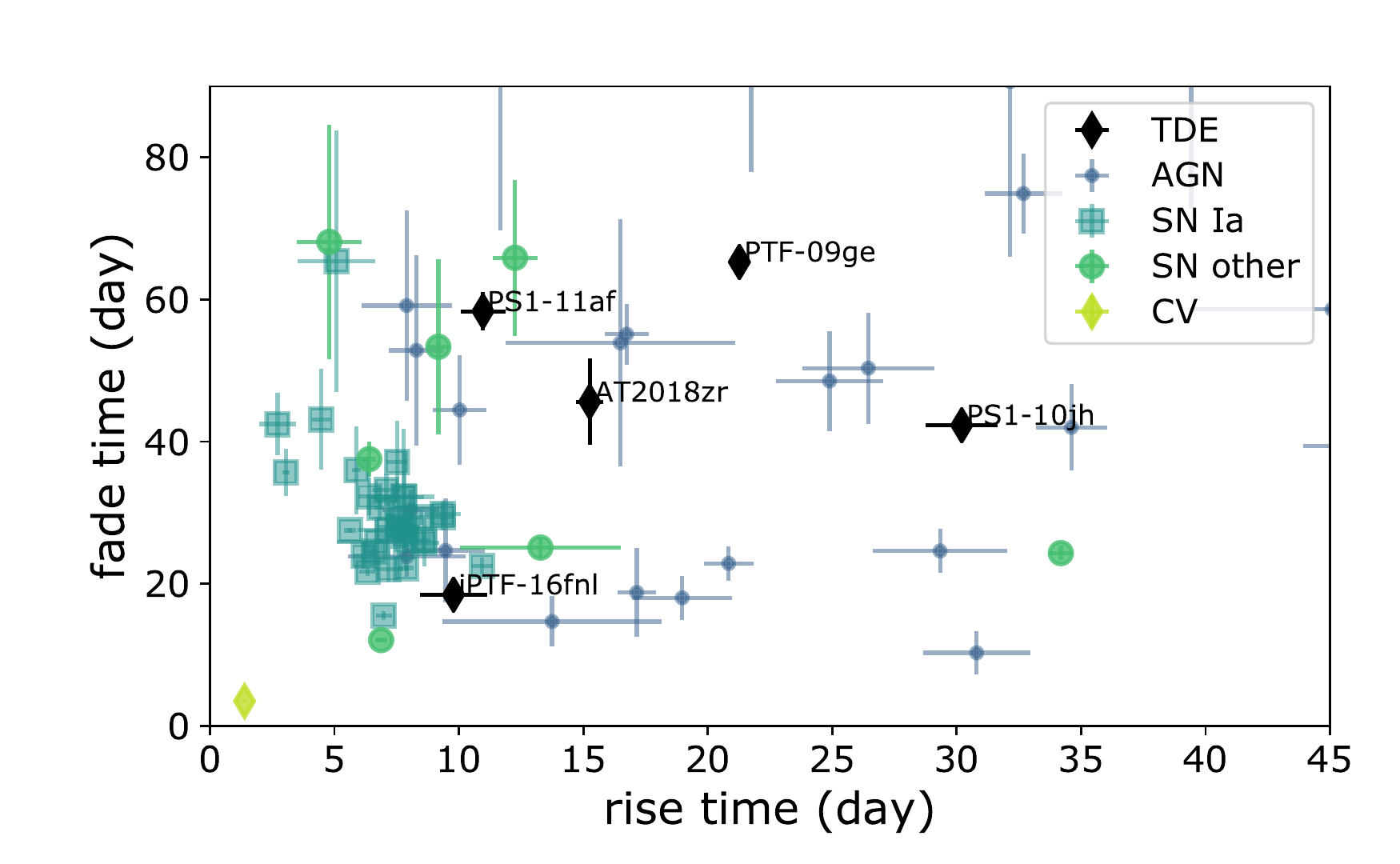}
}
\end{center}
\caption{
{\footnotesize {\it Left panel}: Rate of change of $g-r$ color vs. mean $g-r$ color for nuclear transients from first $\sim$ 6 months of the ZTF Survey.  Note that known TDEs lie in a distinct region of this parameter space, with very blue mean color $g-r < -0.1$ and little color evolution.
{\it Right panel}: Fading timescale vs rise time in days for the same sample of nuclear transients from ZTF.  Note that all of the 5 TDEs discovered before peak have slow rise times relative to SNe Ia, with the exception of the faint and fast evolving iPTF16fnl (Blagorodnova et al. 2017), but still require a cadence of better than 5 days to guarantee a well measured rise to peak.}}
\label{fig1}
\end{figure}

\vspace{.6in}

\section{Technical Description}
\begin{footnotesize}
{\it Describe your survey strategy modifications or proposed observations. Please comment on each observing constraint
below, including the technical motivation behind any constraints. Where relevant, indicate
if the constraint applies to all requested observations or a specific subset. Please note which 
constraints are not relevant or important for your science goals.}
\end{footnotesize}

The metrics that we are focusing on are 1) how many days before peak a TDE is detected, 2) how many points on the pre-peak light curve in g and r, and 3) how many u-band observations are there within +/- 1 week of peak.  The baseline cadence plan (kraken\_2026) have the following attributes in comparison to three other rolling cadence simulations (data from AlternateStrategiesComparison.md (GitHub: oboberg) in terms of their total area covered per year, the median number of visits in each filter, and the median inter-night gap in each filter:

\begin{center}
\begin{tabular}{lccccc}
\hline
 & kraken\_2026 & nexus\_2097 & mothra\_2049 & kraken\_2036 & mothra\_2045\\
 \hline \hline
WFD area/yr (deg$^{2}$) & 18,000 & 6,000 & 9,000 & 6,000 & 9,000\\
median visits (u) & 64 & 46 & 46 & 51 & 47\\
median visits (g) & 90 & 64 & 65 & 71 & 67\\
median visits (r) & 206 & 149 & 153 & 166 & 157\\
inter-night gap (u) & 23.959 & 2.978 & 3.992 & 3.015 & 4.984\\
inter-night gap (g) & 25.858 & 16.952 & 16.029 & 16.928 & 19.891\\
inter-night gap (r) & 7.941 & 3.971 & 3.983 & 3.971 & 3.984\\
\hline
\end{tabular}
\end{center}

It is clear from the table above that the rolling cadences, which alternate strips of 1/2 or 1/3 the WFD survey each year, show a vast improvement in the median inter-night gap in the r band (4 days vs. 8 days in the baseline) and the u band ($3-5$ days vs 24 days).  In order to have a u band observation within +/- 1 week of the peak, for classification purposes, a rolling cadence design is required.  In the baseline configuration (kraken\_2026), the $u$ band data points will be observed monthly, thus delaying the confirmation of a source as having a blue $u-r$ color with no color evolution to over 2 months after the first detection in the $u$ band!  This could potentially be months after peak, and way too late to trigger multi-wavelength follow-up for characterization, and for fainter sources, it is too late since the source will have faded below detection limits already.  Similarly, in the baseline configuration (kraken\_2026) the $g-r$ color of a transient will not be known until the first $g$ band detection, which could be up to a month after peak, and again the $g-r$ color evolution will not be known until the second $g$ band detection, again at least 2 months later!  $u-r$ and $g-r$ color and color evolution are absolutely critical for distinguishing between TDEs and SNe that happen to explode near the center of a galaxy (van Velzen et al. 2011; Hung et al. 2018; van Velzen et al. 2018)

\subsection{High-level description}
\begin{footnotesize}
{\it Describe or illustrate your ideal sequence of observations.}
\end{footnotesize}

To maximize transient science, and the ability to classify transients in a timely manner (or even at all for more rapidly evolving transients), more visits are needed per month over the yearly visibility window of each field.  To achieve this, we are advocating for ``smart" rolling cadence design, with up to only 1000 deg$^{2}$ observed in a given month.  To enable the creation of reference images, and to satisfy early science requirements of other LSST science areas, if the WFD was covered in full in Year 1 and Year 10, and Years 2-9 were dedicated to the rolling survey, then the 20,000 deg$^{2}$ footprint could be divided into 8 strips, each 2500 deg$^{2}$ in area, observed in each year of the rolling survey.\\

\noindent Year 1: Full WFD (20,000 deg$^{2}$)\\
Year 2:  Strip 1 (2500 deg$^{2}$)\\
Year 3:  Strip 2 (2500 deg$^{2}$)\\
Year 4:  Strip 3 (2500 deg$^{2}$)\\
Year 5:  Strip 4 (2500 deg$^{2}$)\\
Year 6:  Strip 5 (2500 deg$^{2}$)\\
Year 7:  Strip 6 (2500 deg$^{2}$)\\
Year 8:  Strip 7 (2500 deg$^{2}$)\\
Year 9:  Strip 8 (2500 deg$^{2}$)\\
Year 10: Full WFD (20,000 deg$^{2}$)\\

Even if of the total 2500 deg$^{2}$ of each strip, only 1000 deg$^{2}$ is visible in any given month, then that area surveyed to 24.5 mag in r band would still yield 200 TDEs per year, but with the guarantee of a well sampled light curve in $u$ and $r$ bands, allowing for early classification and prompt follow-up.  This would be an order of magnitude jump in the 
detection capabilities of ZTF and ASASSN, and guarantee a legacy sample of light curves that can be used on their own to understand the physical parameters of the LSST TDE sample (rise time, peak luminosity, temperature, radius, etc.).   
\vspace{.3in}

\subsection{Footprint -- pointings, regions and/or constraints}
\begin{footnotesize}{\it Describe the specific pointings or general region (RA/Dec, Galactic longitude/latitude or 
Ecliptic longitude/latitude) for the observations. Please describe any additional requirements, especially if there
are no specific constraints on the pointings (e.g. stellar density, galactic dust extinction).}
\end{footnotesize}

The important parameter is area monitored per month in the WFD Survey.  This should be at least 1000 deg$^{2}$.  If the WFD was covered in full in Year 1 and Year 10, and Years 2-9 were dedicated to the rolling survey, then the 20,000 deg$^{2}$ footprint could be divided into 8 strips, each 2500 deg$^{2}$ in area, observed in each year of the rolling survey, corresponding to an active area monitored in any given month of $\sim 1000$ deg$^{2}$.  Even if only 500 deg$^{2}$ WFD strips were observed with this cadence each year, that would still yield a total sample of 50 TDEs per year for which photometrically classified and characterized ensemble studies could be carried out with transformative results.

\subsection{Image quality}
\begin{footnotesize}{\it Constraints on the image quality (seeing).}\end{footnotesize}

\noindent No special requirements.

\subsection{Individual image depth and/or sky brightness}
\begin{footnotesize}{\it Constraints on the sky brightness in each image and/or individual image depth for point sources.
Please differentiate between motivation for a desired sky brightness or individual image depth (as 
calculated for point sources). Please provide sky brightness or image depth constraints per filter.}
\end{footnotesize}

\noindent No special requirements.

\subsection{Co-added image depth and/or total number of visits}
\begin{footnotesize}{\it  Constraints on the total co-added depth and/or total number of visits.
Please differentiate between motivations for a given co-added depth and total number of visits. 
Please provide desired co-added depth and/or total number of visits per filter, if relevant.}
\end{footnotesize}

The total number of observations each month of the WFD rolling strip should be $7-10$ visits in $g$ and $r$ and 6 visits in $u$.

\subsection{Number of visits within a night}
\begin{footnotesize}{\it Constraints on the number of exposures (or visits) in a night, especially if considering sequences of visits.  }
\end{footnotesize}

\noindent No special requirements.

\subsection{Distribution of visits over time}
\begin{footnotesize}{\it Constraints on the timing of visits --- within a night, between nights, between seasons or
between years (which could be relevant for rolling cadence choices in the WideFastDeep. 
Please describe optimum visit timing as well as acceptable limits on visit timing, and options in
case of missed visits (due to weather, etc.). If this timing should include particular sequences
of filters, please describe.}
\end{footnotesize}

Optimum scheduling of visits in a filter would be evenly distributed throughout the observing window for that rolling strip.

\subsection{Filter choice}
\begin{footnotesize}
{\it Please describe any filter constraints not included above.}
\end{footnotesize}

$u$, $g$, and $r$ are the most important filters for TDE detection and classification.

\subsection{Exposure constraints}
\begin{footnotesize}
{\it Describe any constraints on the minimum or maximum exposure time per visit required (or alternatively, saturation limits).
Please comment on any constraints on the number of exposures in a visit.}
\end{footnotesize}

\noindent No special requirements.

\subsection{Other constraints}
\begin{footnotesize}
{\it Any other constraints.}
\end{footnotesize}

\subsection{Estimated time requirement}
\begin{footnotesize}
{\it Approximate total time requested for these observations, using the guidelines available at \url{https://github.com/lsst-pst/survey_strategy_wp}.}
\end{footnotesize}

Same as default baseline WFD survey time requested.

\vspace{.3in}

\begin{table}[ht]
    \centering
    \begin{tabular}{l|l|l|l}
        \toprule
        Properties & Importance \hspace{.3in} \\
        \midrule
        Image quality &   2  \\
        Sky brightness &  2\\
        Individual image depth &  2 \\
        Co-added image depth &  3 \\
        Number of exposures in a visit   &   3\\
        Number of visits (in a night)  &  3 \\ 
        Total number of visits &  1 \\
        Time between visits (in a night) & 3 \\
        Time between visits (between nights)  & 1  \\
        Long-term gaps between visits & 1\\
        Other (Time between visits in different filters) & 1\\
        \bottomrule
    \end{tabular}
    \caption{{\bf Constraint Rankings:} Summary of the relative importance of various survey strategy constraints. Please rank the importance of each of these considerations, from 1=very important, 2=somewhat important, 3=not important. If a given constraint depends on other parameters in the table, but these other parameters are not important in themselves, please only mark the final constraint as important. For example, individual image depth depends on image quality, sky brightness, and number of exposures in a visit; if your science depends on the individual image depth but not directly on the other parameters, individual image depth would be `1' and the other parameters could be marked as `3', giving us the most flexibility when determining the composition of a visit, for example.}
        \label{tab:obs_constraints}
\end{table}

\subsection{Technical trades}
{\it To aid in attempts to combine this proposed survey modification with others, please address the following questions:}
\begin{enumerate}
   {\it \item What is the effect of a trade-off between your requested survey footprint (area) and requested co-added depth or number of visits?}\\
    No trade-off in total area or co-added depth.  They are the same as baseline survey.
   {\it  \item If not requesting a specific timing of visits, what is the effect of a trade-off between the uniformity of observations and the frequency of observations in time? e.g. a `rolling cadence' increases the frequency of visits during a short time period at the cost of fewer visits the rest of the time, making the overall sampling less uniform.}\\
    The extreme rolling cadence proposed in this WP removed the ability to study variability of sources on timescales of greater than 1 year.  However, it does still include the 10 year baseline of Year 1 and Year 2.  Thus for stochastically varying sources, you will get observations in Year 1, one year in Year 2-9, and observations in Year 10.  This should be sufficient to probe timescales of 1 week, 1 month, 6 months, 1-9 years, and 10 years.
   {\it  \item What is the effect of a trade-off on the exposure time and number of visits (e.g. increasing the individual image depth but decreasing the overall number of visits)?}\\
    The default plan of 2 visits per filter on a given night for moving object identification is fine.  However, if more than 2 visits are taken on a night, then the distribution of visits over the month will be worse, and the ability to study transients on 3-5 days timescales will be lost.
    {\it \item What is the effect of a trade-off between uniformity in number of visits and co-added depth? Is there any benefit to real-time exposure time optimization to obtain nearly constant single-visit limiting depth?}\\
    The uniform distribution of visits over the observing window is more important that the limiting depth per visit.
   {\it  \item Are there any other potential trade-offs to consider when attempting to balance this proposal with others which may have similar but slightly different requests?}\\
     The key question is what science collaborations require time domain observations on longer than a timescale of 6 months...and if they do require longer timescales, how evenly do those observations have to be distributed, and how much area do they really need.  For example, for TDE science, we would rather have 100 well sampled TDE light curves per year from LSST, instead of 4000 per year that we cannot classify, or use the LSST light curves to measure any parameters about the events.
\end{enumerate}

\section{Performance Evaluation}
\begin{footnotesize}
{\it Please describe how to evaluate the performance of a given survey in achieving your desired
science goals, ideally as a heuristic tied directly to the observing strategy (e.g. number of visits obtained
within a window of time with a specified set of filters) with a clear link to the resulting effect on science.
More complex metrics which more directly evaluate science output (e.g. number of eclipsing binaries successfully
identified as a result of a given survey) are also encouraged, preferably as a secondary metric.
If possible, provide threshold values for these metrics at which point your proposed science would be unsuccessful 
and where it reaches an ideal goal, or explain why this is not possible to quantify. While not necessary, 
if you have already transformed this into a MAF metric, please add a link to the code (or a PR to 
\href{https://github.com/lsst-nonproject/sims_maf_contrib}{sims\_maf\_contrib}) in addition to the text description. (Limit: 2 pages).}
\end{footnotesize}

This "smart" and "colorful" rolling cadence plan, with 2500 deg$^{2}$ WFD strips observed each year, with 1000 deg$^{2}$ of sky observed any given month, would yield 200 TDEs per year with well sampled light curves on the rise to peak, and well measured $u-r$ and $g-r$ color and color evolution to allow for efficient selection from the background of much more common transients such as SNe and flaring AGN.  We plan to write software to return a figure of merit for TDE/SNe discrimination; this will be based on the ratio of SNe and TDE in a given box in color space (cf. figure 2) and the total number of TDEs that can identified using these boxes.  The rise to peak in a TDE is related to the fallback timescale of the tidally disrupted star, which scales with the central black hole mass as $t_{\rm fb} = 41 $d $(M_{\rm BH}/(10^6 M_\odot)^{1/2}$ (Lodato et al. 2009; Guillochon \& Ramirez-Ruiz 2013) .  Thus for the range of SMBHs from which we expect to detect TDEs by LSST ($10^5-10^8 M_\odot$), we need a cadence of at most 5 days to probe that rise in detail, and constrain $M_{\rm BH}$)  This is also known observationally, where the observed rise to peak timescale in the only 5 TDEs with pre-peak light curves ranges from as short at 10 days (iPTF-16fnl) to 30 days (PS1-10jh) (see Figure below from van Velzen et al. 2018).  {\bf To conclude, we are confident that for LSST to produce a legacy sample of light curves that are fruitful for TDE science (as well as SN studies), a denser multi-band cadence must be performed over a fraction of the WFD survey area that is much larger (at least by a factor of 10) than the current plan for the Deep Drilling Fields. } 
 
\vspace{.6in}

\section{Special Data Processing}
\begin{footnotesize}
{\it Describe any data processing requirements beyond the standard LSST Data Management pipelines and how these will be achieved.}
\end{footnotesize}

No special requirements.

\section{Acknowledgements}
This work developed partly within the LSST Transient and Variable Stars (TVS) Science Collaboration and the author acknowledge the support of TVS in the preparation of this paper.  S.G. acknowledge support from the Flatiron Institute and Heising-Simons Foundation for the development of this paper.

\section{References}

Blagorondova et al. 2017, ApJ, 844, 46\\
Gezari, S. et al. 2009, ApJ, 698, 1367 \\
Gezari, S. et al. 2012,Nature, 485, 217 \\
Guillochon, J. \& Ramirez-Ruiz, E. 2013, ApJ, 767, 25 \\
Holoien, T. et al. 2014, MNRAS, 445, 326 \\
 Holoien, T. et al. 2016, MNRAS, 463, 3813 \\
 Hung, T. et al. 2018, ApJs, 238, 15) \\
 Lodato, G. et al. 2009, MNRAS, 392, 332 \\
 Miller, J. et al. 2015, Nature, 526, 542 \\
 Mockler, B. et al. 2018, ApJ, submitted (arXiv:1801.08221) \\
 Pasham, D. et al. 2017, ApJ, 837, L30 \\
 van Velzen, S., et al. 2018, ApJ, submitted (arXiv:1809.02608)\\

\end{document}